\documentclass{ifacconf}

\usepackage[utf8]{inputenc}
\usepackage{amsfonts}
\usepackage{booktabs}
\usepackage{mathrsfs}
\usepackage{amsmath} % assumes amsmath package installed
\usepackage{amssymb}  % assumes amsmath package installed
\usepackage{natbib}            % you should have natbib.sty
\usepackage{graphicx}          % Include this line if your
\usepackage{epstopdf}
\usepackage{subfigure}

\usepackage[dvips]{epsfig}    % or this line, depending on which
\usepackage{algorithm}
\usepackage[noend]{algpseudocode}

\makeatletter
\def\BState{\State\hskip-\ALG@thistlm}
\makeatother

\allowdisplaybreaks

\newif\ifshowestimation

\showestimationfalse

\newenvironment{talign}
 {\let\displaystyle\textstyle\align}
 {\endalign}
\newenvironment{talign*}
 {\let\displaystyle\textstyle\csname align*\endcsname}
 {\endalign}

\begin{document}
\begin{frontmatter}

\title{The role of noise modeling in the estimation of resting-state brain effective connectivity} 

\thanks{This work has been partially supported by the  project BIRD162411/16   and the FIRB project  RBFR12M3AC funded by MIUR.}

\author[First]{G. Prando} 
\author[First]{M. Zorzi} 
\author[First]{A. Bertoldo} 
\author[First]{A. Chiuso}

\address[First]{Dept. of Information  Engineering, University of Padova (e-mail: \{prandogi,zorzimat,bertoldo,chiuso\}@dei.unipd.it)}

\begin{abstract}                
Causal relations among  neuronal populations of the brain are studied through the so-called effective connectivity (EC) network. The latter is estimated from EEG or fMRI measurements, by inverting a generative model of the corresponding data. It is clear that the goodness of the estimated network heavily depends on the underlying modeling assumptions. In this present paper we consider the EC estimation problem using fMRI data in resting-state condition. Specifically, we investigate on how to model endogenous fluctuations driving the neuronal activity. 
\end{abstract}

\begin{keyword}
System identification,  Estimation and filtering, Neuroscience \end{keyword}

\end{frontmatter}
%===============================================================================

\section{Introduction} \label{sec:intro}
A deeper understanding of the complex interactions which take place inside our brain, which  currently represents a major open research challenge,  could possibly open new avenues to novel terapeutic treatments as well as early diagnosis tools.  It is thus not surprising that this area is attracting the interest of scientists from many different fields, as it calls for diverse expertise ranging from medical science to physics, from biology to engineering and statistics. In particular, the estimation of effective connectivity, accounting for the causal relationships among different neuronal populations, has been the subject of several research contributions in the last decade \citep{valdes2011effective,razi2016connected}.

Effective connectivity is typically inferred from EEG or fMRI measurements, by inverting a generative model of the corresponding data. With the language of System Identification, estimating an effective connectivity model can be framed as the problem of inferring a dynamic network model from indirect measurements. In fact, such models should not only describe the causal interactions taking place at neuronal level (described by the so-called effective connectivity), but also capture the mapping between such activity and the available measurements. Specifically, when the fMRI technique is adopted, this mapping is called hemodynamic response and links the underlying synaptic dynamics to the so-called BOLD (Blood Oxygenation Level Dependent) signal, which is measured through fMRI. This signal reflects an increase of the neuronal activity on a change of the relative levels of oxyhemoglobin and deoxyhemoglobin.

%Indeed, both EEG and fMRI only provide indirect observations of the underlying neuronal activity. Specifically, fMRI detects the so-called BOLD (Blood Oxygenation Level Dependent) signal, which measures changes of the relative levels of oxyhemoglobin and deoxyhemoglobin. The relationship between the BOLD signal and the underlying synaptic activity is regulated by the so-called hemodynamic response. 

The neuroscience literature has proposed different characterizations of the aforementioned generative models, ranging from linear to non-linear models, from discrete-time (DT) to continuous-time (CT) frameworks (see \cite{valdes2011effective} for a comprehensive review). In this work we consider the Dynamical Causal Model (DCM) proposed by \cite{friston2003dynamic}, which is nowadays a widespread approach in the neuroscience community \citep{daunizeau2011dynamic}. A DCM is a CT nonlinear state-space model, where the hemodynamic response is described by the so-called Balloon-Windkessel model \citep{buxton1998dynamics}. Whereas the original version of DCM was suited for fMRI data recorded while the subject is performing a task \citep{friston2003dynamic}, the DCM has also been recently extended to account for resting-state data \citep{friston2014dcm}. While in task-dependent DCM, neuronal dynamics is driven by deterministic inputs (external stimuli), in resting-state DCM it is elicited by stochastic sources, representing brain endogenous fluctuations.  In the present paper, we will focus on DCM for resting-state data.
%\cite{friston2014dcm}, as well as \cite{razi2015construct}, postulate a power-law spectrum for these random components, thus accounting for the scale-free properties of neuronal fluctuations which have been observed in previous works . Thus, according to such modeling choice, the spectrum of the neuronal noise is concentrated at low-frequency.

In \cite{friston2014dcm} and \cite{razi2015construct}, brain endogenous fluctuations are assumed to have a power-law spectrum. Such choice not only leads to a low frequency-concentrated spectrum, but also accounts for the scale-free properties of neuronal fluctuations which have been observed in previous works \citep{stam2004scale,linkenkaer2001long,shin2006self}. Other contributions by the same authors \citep{friston2008hierarchical,friston2008variational} suggest the use of non-Markovian noise processes when modeling biophysical fluctuations, such as those taking place in the brain.

In our previous work \citep{prando2017estimating} we tackled the estimation of resting-state effective connectivity by adopting a linearized DCM and postulating that neuronal activity is driven by white Gaussian  noise; the effective connectivity matrix was inferred by imposing a sparsity inducing prior \citep{wipf2010iterative}. The remarkable advantage of such a choice is that we are able to avoid the combinatorial bottleneck which typically characterizes the DCM inversion algorithms. The main contribution of this work is twofold: first, following the modeling assumptions for the endogenous fluctuations suggested in the aforementioned literature,  we extend the model used in \cite{prando2017estimating} allowing for a CT autoregressive (AR) model of the random fluctuations driving the neuronal activity. Second,   using   both real and synthetic data, we compare white vs. AR(1) noise models as sources for   the resting-state neuronal activity. This comparison is carried out both studying the effect of noise models in capturing functional correlations as well as evaluating the difference between effective connectivity estimated under different modeling assumptions. Our preliminary results suggest that, indeed, a low-pass AR(1) noise is more suited in our setup.

The manuscript is organized as follows. Modeling aspects are treated in Sec.~\ref{sec:model}, with a specific focus on the modeling of brain endogenous fluctuations. %Sec.~\ref{sec:estimation} briefly illustrates the algorithm we adopt to estimate effective connectivity,
An extensive simulation study is reported in Sec.~\ref{sec:experiments}. Concluding remarks and future investigations are outlined in Sec.~\ref{sec:conclusions}.

{\bf Notation}:   $I_n$ denotes the identity matrix of size $n$ and  $0_{m\times n}$ the $m\times n$ zero matrix;  $\mathrm{(blk)diag(\cdot)}$ denotes the (block-)diagonal operator. $A^T$ indicates the tranpose of matrix $A$.

\section{Effective Connectivity Modeling}\label{sec:model}
Inference of brain effective connectivity requires the definition of a generative model of the observed BOLD time-series measured by fMRI scanners. 
Here we use the DCMs  introduced in \citep{friston2003dynamic,friston2014dcm}; these are   non-linear state-space models which jointly describe 
the neuronal activity and the map from the neural activity to  the BOLD signal. Denoting with $x(t) =[x_1(t),...,x_n(t)]^T\in\mathbb{R}^n$ the activity of $n$ 
neuronal populations at time $t\in\mathbb{R}$, its evolution is described as
\begin{equation}\label{equ:rs_neuro_dyn}
\dot{x}(t) = Ax(t) + w(t)
\end{equation}
where $A\in\mathbb{R}^{n\times n}$ is the effective connectivity matrix. The process $w(t)$ represents random fluctuations, whose modeling is the main focus of this work. Previous contributions \citep{friston2014dcm,razi2015construct} have postulated that $w(t)$ has a power-law power spectrum:
\begin{equation}
W(\omega; (\nu,\beta)) = \nu \omega^{-\beta} 
\end{equation}
with $\nu$ and $\beta$ being parameters to be inferred and $\omega$ denoting the angular frequency.\\
The DCM specification is completed by the so-called Balloon-Windkessel model \citep{buxton1998dynamics}, a 4th-order non-linear SISO system which takes as input each $x_i(t)$ and outputs the measured BOLD signal $y_i(t)$:
\begin{talign}
\dot{r}_i(t) &= x_i(t) - \kappa_i r_i(t) -\gamma_i (f_i(t)-1), \quad i=1,...,n \label{equ:states_balloon_1}\\
\dot{f}_i(t) &= r_i(t)\nonumber\\
\tau_i \dot{v}_i(t) &= f_i(t) - v_i^{1/\xi}(t)\nonumber\\
\tau_i \dot{q}_i(t) &= (f_i(t)/\rho_i) \left[1- (1-\rho_i)^{1/f_i(t)}\right] - v_i^{1/\xi-1}(t)q_i(t)\nonumber \\
y_i(t) &= V_0 k_1 (1-q_i(t)) + V_0k_2 \left(1- q_i(t)/v_i(t)\right)\nonumber\\
& \hspace{0.5cm} + V_0k_3 (1-v_i(t)) + e_i(t).\label{equ:output_balloon}
\end{talign}
The model depends on the parameters $\theta_b := \{\kappa_i,\gamma_i,\tau_i,\rho_i,\xi\}$ that are characterized in the literature by empirical priors and have to be estimated from data \citep{friston2003dynamic}. The states $\{r_i,f_i,v_i,q_i\}$ are biophysical quantities that are affected by the neuronal activity: $r_i$ denotes the vasodilatatory signal, $f_i$ is the blood inflow, $v_i$ and $q_i$ are respectively the blood volume and the deoxyhemoglobin content. The output equation \eqref{equ:output_balloon} includes the measurement noise $e_i$, while the signal component depends on the resting blood volume fraction $V_0$ (typically $V_0=0.4$) and on the constants $k_1$, $k_2$ and $k_3$. 
We refer the reader to \cite{stephan2007comparing} for more details.

In our previous work \citep{prando2017estimating} we proposed an alternative formulation of the DCM
in Eqs.~\eqref{equ:rs_neuro_dyn}-\eqref{equ:output_balloon}. First, since fMRI data are acquired with a sampling time $T_R=$ 2 s., we adopt a sampled version of Eq.~\eqref{equ:rs_neuro_dyn} and model $w(t)$ as a CT white noise with intensity $\sigma I_n$. Then, defining  $x(k):=x(kT_R)$, we discretize Eq.~\eqref{equ:rs_neuro_dyn} as
\begin{equation}\label{equ:discrete_neuro_dyn}
x(k+1) = e^{AT_R}x(k) + w_d(k), \qquad k\in\mathbb{Z}
\end{equation} 
where the noise process $w_d(k)$ satisfies:
\begin{align}
\textstyle{w_d(k)}&:=\textstyle{w_d(kT_R) = \int_0^{T_R} e^{A\tau} w((k+1)T_R - \tau) d\tau}\nonumber\\
\textstyle{{\rm Var}\{w_d(k)\}}  &=   \textstyle{Q_w = \sigma \int_0^{T_R} e^{A\tau} e^{A^T \tau} d\tau.} \label{equ:discrete_noise}
\end{align}
%Notice that  Eq.~\eqref{equ:discrete_neuro_dyn} adopts the simplified notation $x(k):=x(kT_R)$ and $w_d(k):=w_d(kT_R)$.\\
Second, we have  proposed a statistical linearization of the Balloon-Windkessel model, exploiting the empirical priors for the hemodynamic parameters $\theta_b$. Specifically, we have reformulated model (\ref{equ:states_balloon_1})-(\ref{equ:output_balloon}) as an FIR model
\begin{equation}\label{equ:fir_hemo}
\textstyle{b_i(k) =  \sum_{l=0}^{s-1}\ [h_i]_l\ x_i(k-l)}, \qquad i=1,...,n
\end{equation}
with a prior on $h_i\in\mathbb{R}^s$ derived by the prior on  $\theta_b$. The length $s$ of the impulse response $h$ is chosen large enough to capture all the relevant dynamics (see Sec.~III in \cite{prando2017estimating}). 
Using the linearisation \eqref{equ:fir_hemo} and defining 
\begin{align}\label{equ:a_c}
\mathbf{C} :=& \begin{bmatrix}
\mathrm{diag}([h_1]_0, ..., [h_n]_0) | \cdots | \mathrm{diag}([h_1]_{s-1}, ..., [h_n]_{s-1})
\end{bmatrix},\nonumber\\
\mathbf{A} :=&  \begin{bmatrix}
e^{AT_R} &  0_{n\times n(s-1)}\\ I_{n(s-1)} & 0_{n(s-1)\times n}
\end{bmatrix},\\
%\mathbf{C} &:= \left[\begin{array}{ccc|c|ccc}
%[h_1]_0 & \cdots  & 0 & \cdots & [h_1]_{s-1} & 0 & 0\\
%\vdots & \ddots & \vdots & \cdots & 0 &\ddots & \vdots \\
%0 & \cdots & [h_n]_0 & \cdots & 0 & \cdots & [h_n]_{s-1}
%\end{array}\right]\\
\mathbf{x}(k):=& \begin{bmatrix}
x^T(k) & x^T(k-1) & \cdots & x^T(k-s+1)
\end{bmatrix}^T,\\
\mathbf{w}_d(k):&=[w_d^T(k)\ 0_{1 \times n(s-1)}]^T, \\
\mathbf{Q}_w:=&\mathrm{Var}\{\mathbf{w}_d(k)\}=\mathrm{blkdiag}(Q_w,0_{n(s-1)\times n (s-1)}),\nonumber
\end{align} 
the resting state  DCM becomes  a linear stochastic state-space model; namely:
\begin{equation}\label{equ:linear_stochastic_dcm}
\left\{\begin{array}{rcl}
\mathbf{x}(k+1) & = & \mathbf{A} \mathbf{x}(k) + \mathbf{w}_d(k) \\
y(k) &=& \mathbf{C} \mathbf{x}(k) + e(k)
\end{array}\right.
\end{equation}

Exploiting the linearity of model \eqref{equ:linear_stochastic_dcm} as well as a suitable  sparsity inducing prior on $A$, in \cite{prando2017estimating} we have introduced an EM-type iterative procedure which is much less computationally demanding than  variational methods typically employed to invert classical DCMs.  For reasons of space we refer the interested reader to \cite{prando2017estimating} for details on the estimation algorithm.

Choosing the most suitable  statistical description of  the endogenous fluctuations $w(t)$  in Eqs.~\eqref{equ:rs_neuro_dyn} and \eqref{equ:discrete_neuro_dyn} is subject of current research and debate in the scientific community. The simplest choice, which we have followed in \cite{prando2017estimating}, is to model  $w(t)$  as CT white noise. Instead, motivated by the findings of \cite{linkenkaer2001long,stam2004scale},  \cite{friston2014dcm} and \cite{razi2015construct} postulate that the power spectrum of $w(t)$ is concentrated at low frequencies. Following this line of work, in this paper we 
reformulate system \eqref{equ:linear_stochastic_dcm} assuming $w(t)$ to be a 1st-order CT autoregressive process.

\subsection{Vector Autoregressive (VAR) Process Noise}\label{subsec:ar}
According to the discussion in the previous section, we could either model $w(t)$ in \eqref{equ:rs_neuro_dyn} as a CT AR process, or  we could 
directly consider the discrete time counterpart Eq.~\eqref{equ:discrete_neuro_dyn} and model $w_d(k)$
as  DT coloured noise. Following the first option we model $w(t)$ using the CT model \begin{equation}\label{equ:ct_ar}
\dot{w}(t) = \Lambda w(t) + v(t)
\end{equation}
where $v(t)$ is a CT white noise with intensity $\delta I_n$ and $\Lambda\in\mathbb{R}^{n\times n}$ is an Hurwitz-stable matrix. Alternatively, $w_d(k)$ can be described by the
 discrete-time model
\begin{equation}\label{equ:dt_ar}
w_d(k+1) = \Delta w_d(k) + \tilde{v}_d(k)
\end{equation}
with $\Delta\in\mathbb{R}^{n\times n}$ Schur-stable matrix and $\tilde v_d(k)$ a DT white noise with covariance $\widetilde{Q}_v$. These two options lead in general to different models. However, in the special case  $\Lambda=\lambda I_n$, the models in  \eqref{equ:ct_ar}-\eqref{equ:dt_ar} are equivalent under a proper choice of $\Delta$ and $\tilde Q_v$. 
Indeed, sampling Eq.~\eqref{equ:ct_ar}, we get
\begin{talign} \label{eq_w_incr_MZ}w(( &k+1)T_R)\nonumber\\ &= e^{\Lambda T_R} w(kT_R) + \int_0^{T_R} e^{\Lambda s} v((k+1)T_R -s)  ds.\end{talign}
Using (\ref{eq_w_incr_MZ}), we have 
\begin{talign}
\int_0^{T_R} e^{A\tau} &w((k+1)T_R-\tau)d\tau
=\label{equ:dt_ar_1}\\ &\int_0^{T_R} e^{A\tau} e^{\Lambda T_R} w(kT_R-\tau)d\tau\nonumber\\
&+ \int_{0}^{T_R} \int_{0}^{T_R} e^{A \tau} e^{\Lambda s} v((k+1)T_R-s-\tau) ds d\tau.\nonumber
\end{talign}
Now, assuming $\Lambda = \lambda I_n$ and recalling Eq.~\eqref{equ:discrete_noise}, we can rewrite Eq.~\eqref{equ:dt_ar_1} as the DT AR model
\begin{equation}\label{equ:dt_ar_2}
w_d(k) = e^{\lambda T_R} w_d(k-1) + v_d(k) 
\end{equation}
where $v_d(k)$ has covariance 
\begin{equation}
\textstyle{Q_v = \sigma \int_{0}^{T_R} \int_{0}^{T_R} e^{A \tau} e^{2\lambda s} e^{A^T \tau} ds d\tau}
\end{equation}
which directly depends on the dynamics of model \eqref{equ:rs_neuro_dyn}. Hence, the two DT AR models in Eqs.~\eqref{equ:dt_ar} and \eqref{equ:dt_ar_2} coincide only if $\Delta=e^{\lambda T_R}$ and $\tilde{Q}_v$ is set equal to $Q_v$. 

Given  previous neuroscience studies \citep{stam2004scale} and the fact that  $w(t)$ directly represent the brain endogenous fluctuations, in the following we shall work with
the AR(1) CT noise model \eqref{equ:ct_ar} where $\Lambda = \mathrm{diag}(\lambda_1,...,\lambda_n)$, $\lambda_i\in\mathbb{R}_-$. Under this hypothesis Eq.~\eqref{equ:rs_neuro_dyn} can be rewritten as
\begin{equation}\label{equ:w_x}
\begin{bmatrix}
\dot{w}(t) \\ \dot{x}(t) 
\end{bmatrix} = \begin{bmatrix}
\Lambda & 0_n \\ I_n &  A 
\end{bmatrix}\begin{bmatrix}
w(t) \\ x(t) 
\end{bmatrix} + \begin{bmatrix}
v(t) \\ 0_{n\times 1}
\end{bmatrix} =: M \begin{bmatrix}
w(t) \\ x(t) 
\end{bmatrix}+ \eta(t).
\end{equation}
where the last equation defines $M\in\mathbb{R}^{2n\times 2n}$ and $\eta(t)\in\mathbb{R}^{2n}$.
The CT model \eqref{equ:w_x} can be discretised  obtaining:
\begin{equation}\label{equ:w_x_discrete}
\begin{bmatrix}
w(k+1) \\ x(k+1) 
\end{bmatrix} = e^{M T_R}\begin{bmatrix}
w(k) \\ x(k) 
\end{bmatrix} + \eta_d(k)
\end{equation}
where, again, the notation $x(k):=x(kT_R)$ has been used (analogously for $w(k)$ and $\eta_d(k)$) and $\eta_d(k)$ is a stationary white noise with covariance
\begin{equation}
\textstyle{
Q_\eta = \int_0^{T_R} e^{M\tau} \Sigma e^{M^T\tau}d\tau, \quad \Sigma =\mathrm{blkdiag}(0_{n\times n}, \delta I_n)
}
\end{equation} 
Using the statistical linearization of the Balloon-Windkessel model as above,   the linear stochastic state space model
\begin{equation}\label{equ:linear_stochastic_dcm_ext}
\left\{\begin{array}{rcl}
\begin{bmatrix}
w(k+1) \\ \mathbf{x}(k+1)
\end{bmatrix} & = & \mathbf{M} \begin{bmatrix}
w(k) \\ \mathbf{x}(k)
\end{bmatrix} + \boldsymbol{\eta}_d(k) \\
y(k) &=& \mathbf{C} \mathbf{x}(k) + e(k),
\end{array}\right.
\end{equation}
is obtained where, letting $n':= n(s-2)$,
%\begin{equation}
%\mathbf{A}_e :=  \begin{bmatrix}
%e^{A_eT_R} &  0_{2n\times n(s-1)}\\ 0_{n(s-1)\times n} &  I_{n(s-1)} &  0_{n(s-1)\times n}
%\end{bmatrix}
%\end{equation}
\begin{equation}
\mathbf{M} :=  \left[\begin{array}{cc|cc}
\multicolumn{2}{c|}{e^{M T_R}} &  0_{2n\times n'} & 0_{2n\times n}\\
\hline
0_{n} & I_n & 0_{n\times n'} & 0_{n}\\
% \multicolumn{2}{c|}{0_{n' \times 2n}} &  I_{n'} &  0_{n'\times n}
0_{n' \times n} & 0_{n'\times n} &  I_{n'} &  0_{n'\times n}
\end{array}\right]
\end{equation}
The process noise $\boldsymbol{\eta}_d(k):=[\eta_d^T(k)\ 0_{1\times n(s-1)}]^T$ is white with covariance $\mathbf{Q}_\eta:=\mathrm{blkdiag}(Q_\eta,0_{n(s-1)\times n (s-1)})$.
%\begin{equation}
%\mathbf{Q}_\eta:=\mathrm{blkdiag}(Q_\eta, \varepsilon I_{n(s-1)}).
%\end{equation}
It should be noticed that the order of model \eqref{equ:linear_stochastic_dcm_ext} has increased only by $n$, w.r.t. that of \eqref{equ:linear_stochastic_dcm}.
The algorithm in \cite{prando2017estimating}  can also be applied to this scenario with obvious modifications. We thus refer the reader to \cite{prando2017estimating}  for details. 
\begin{rem}
We should stress that modeling $w(t)$ as an AR(1) process coincides with a specific instance of the hierarchical model proposed by \cite{friston2008hierarchical} (see Eqs.~(2)-(3)), which arises when the dynamics is expressed only in terms of the first two generalized coordinates of motion.
\end{rem}

\section{Experiments}\label{sec:experiments}
This section aims at validating (or invalidating) the assumption that the endogenous  fluctuations should be modeled as an  autoregressive process. To this purpose we shall perform experiments on both synthetic as well as real data. Concerning the synthetic experiment,  we shall simulate fMRI signals exciting the simple brain model \eqref{equ:rs_neuro_dyn}-\eqref{equ:output_balloon} with endogenous brain fluctuations $w(t)$ which are either white  or AR(1) noise. Both for simulated and real data we estimate the effective connectivity (EC), using either model \eqref{equ:linear_stochastic_dcm} and \eqref{equ:linear_stochastic_dcm_ext} (assuming $\Lambda$  diagonal). We shall then compare the results with the goal of understanding which modelling assumption is most suited to describe the real fMRI data.

\subsection{Synthetic Data.}
Synhetic neural responses are simulated using the discretizations \eqref{equ:discrete_neuro_dyn} and \eqref{equ:w_x_discrete}  with $T_R=0.05$s, with the ``true'' EC matrix given by
 \begin{equation}\label{equ:true_A}
\footnotesize A = \begin{bmatrix}
-0.5 & 0 & 0 & 0 & -0.2 & 0 & 0\\
0 & -0.5 & 0 & -0.45 & -0.3 & 0 & 0\\
0 & 0 & -0.5 & 0.8 & 0 & 0 & 0\\
0 & 0.6 & 0 & -0.5 & -0.1 & 0.6 &0\\
0.3 & 0 & -0.55 & 0 & -0.5 & 0.2 & 0\\
0 & 0 & 0 & 0 & 0.3 & -0.5 & 0.45\\
0.15 & 0 & 0.2 & 0 & 0 & 0 & -0.5
\end{bmatrix}
\end{equation} 
The BOLD signal is then obtained giving these simulated neural responses as inputs to the Balloon-Windkessel model \eqref{equ:states_balloon_1}-\eqref{equ:output_balloon}. The output is then down-sampled to $T_R=2$s (the sampling rate commonly used in fMRI scanners), obtaining estimation ($\mathcal{D}_i$)  and test ($\mathcal{D}^{te}_i$) datasets , $i=1,..,50$, each containing a BOLD time-series of length $N=300$. 
%These are generated using different realizations of the noises $w_d(k)$ and $\eta_d(k)$, respectively. In addition, f
For each dataset the Balloon model parameters $\theta_b$ are drawn from their empirical priors reported in \cite{friston2003dynamic} and the measurement noise variance (see Eq.~\eqref{equ:output_balloon}) is chosen so as to guarantee  SNR=10.
% the measurement noise $e(t)$ in Eq.~\eqref{equ:output_balloon} is sampled in order to have SNR=10. 
When using AR input noise (see \eqref{equ:w_x_discrete}) the coefficients $\Lambda$ are drawn for each Monte-Carlo run from a uniform distribution ${\cal U}(-1,0)$. 

\subsection{Real Data.}
Real BOLD time-series from 333 brain regions (ROIs) are measured with $T_R=2$s on two subjects  at rest on 10 minutes scanning sessions, corresponding to BOLD time-series of length $N=300$. \\
The 333 ROIs are then clustered in so-called functional networks. Here we focus on the Visual Network (VIS), responsible for the human vision and on the Default Mode Network (DMN), which collects the most active areas during resting-state condition. In Sec.~\ref{sec:results} we will use the BOLD time-series in VIS and DMN (denoted with $\mathcal{D}^{\mathrm{VIS}}$ and $\mathcal{D}^{\mathrm{DMN}}$ respectively) to infer the EC restricted to these networks. 
% In Sec.~\ref{sec:results} we will illustrate the estimation of the EC restricted to either the Visual Network or to the DMN. 
Finally, the 333 BOLD  time series are reduced to 53 components by averaging  on functionally homogenous areas, obtaining the dataset $\mathcal{D}^{wb}$ which is used to infer the whole-brain EC.

\subsection{Modeling Assumptions}\label{sec:model_assumptions}
We estimate the EC matrix $A$ under three different modeling assumptions (denoted below with $\mathcal{I}:=\{\mathrm{W,AR,VAR}\}$); in brackets the  corresponding symbol used for the estimated EC:
\begin{itemize}
\item $w(t)$ White noise ($\hat{A}_\mathrm{W}$)
\item $w(t)$ AR(1) noise with $\Lambda=\lambda I_n$ ($\hat{A}_\mathrm{AR}$) 
\item $w(t)$ AR(1) noise with $\Lambda=\mathrm{diag}(\lambda_1,...,\lambda_n)$ ($\hat{A}_\mathrm{VAR}$)
\end{itemize}
% In the following we will use the notation $\mathcal{I}:=\{\mathrm{W,AR,VAR}\}$ to refer to these three modeling assumptions.
%We also postulate a Gaussian prior for $[\lambda_1,...,\lambda_n]^T$ ($\lambda$) with mean -$0.5I_n$ (-$0.5$) and covariance $0.5I_n$ (0.5), denoting with $\hat{A}_{\mathrm{RVAR}}$ ($\hat{A}_{\mathrm{RAR}}$) the resulting estimate. 

\subsection{Evaluation Metrics}\label{sec:metrics}
The performance  are evaluated according to two  metrics: one directly considers the estimated EC, while the other involves an indirect measure based on the so-called functional connectivity (FC). As we shall see below, FC is nothing but the correlation between BOLD signals measured in different regions, and thus is a measure of statistical dependencies among brain regions.
%; FC can be derived from EC, while the opposite is not true, since EC can be viewed as the parameter of a model which tries to explain FC. 
 The interested reader is referred to \cite{razi2016connected} for further details on the distinction and relation between EC and FC.

\subsubsection{Effective Connectivity Metrics.} In the synthetic case, where ground-truth EC is available,  we can evaluate
how close the estimated EC is to the ``true'' one using:
\begin{itemize}
\item The \textit{Root-Mean-Squared-Error} (RMSE)
\begin{equation}\label{equ:rmse}
\mathrm{RMSE}(\hat{A}_\mathrm{I}) := \|\underline{A} - \underline{\hat{A}}_{\mathrm{I}}\|_F\ /\, \sqrt{n(n-1)}, \quad \mathrm{I}\in\mathcal{I}
\end{equation}
where the notation $\underline{A}$ denotes the matrix $A$ with its diagonal set to 0\footnote{The diagonal elements of $A$ (i.e. the self-connections)  do not give any useful information on EC.}.
\item The number of errors on the  sparsity pattern of the EC matrix, that is
\begin{equation}\label{equ:err}
\mathrm{ERR}(\hat{A}_\mathrm{I}) := \|SP(\underline{A})-SP(\underline{\hat{A}}_\mathrm{I})\|_F^2 
\end{equation}
where $SP(\cdot): \mathbb{R}^{n\times m} \rightarrow \mathbb{R}^{n\times m}$, $[SP(A)]_{ij}=1$ if $[A]_{ij}\neq 0$, $[SP(A)]_{ij}=0$ otherwise. 
\end{itemize}
On real data, no ground-truth is available and thus the two metrics above cannot be computed. Instead, we can   quantify the dissimilarity arising from the different modeling assumptions listed in Sec.~\ref{sec:model_assumptions} by means of the Pearson's correlation coefficient among the estimated ECs. For two generic vectors $u\in\mathbb{R}^m$ and $v\in\mathbb{R}^m$, this is defined as
\begin{equation}\label{equ:rho}
\rho(u,v) := \frac{\sum_{i=1}^m \left(u_i-\bar{u}\right)\left(v_i-\bar{v}\right)}{\sqrt{\sum_{i=1}^m \left(u_i-\bar{u}\right)^2}\sqrt{\sum_{i=1}^m \left(v_i-\bar{v}\right)^2}}
\end{equation}
where $\bar{u}=\frac{1}{m}\sum_{i=1}^mu_i$ and analogously for $\bar{v}$.\\
Hence, for each pair of estimates within the set $\{\hat{A}_{\mathrm{W}},\hat{A}_{\mathrm{AR}},$ $\hat{A}_{\mathrm{VAR}}\}$ we compute
\begin{equation}
\rho_{\mathrm{EC}}(\mathrm{I},\mathrm{J}):= \rho(\underline{\hat{a}}_\mathrm{I},\underline{\hat{a}}_\mathrm{J}), \qquad \mathrm{I,J}\in\mathcal{I}
\end{equation}
%\begin{equation}
%\rho_{EC}(\mathrm{I},\mathrm{J}):= \frac{\sum_{i=1}^{n(n-1)} ([\underline{\hat{a}}_{\mathrm{I}}]_i-\bar{\hat{\underline{a}}}_{\mathrm{I}})([\underline{\hat{a}}_{\mathrm{J}}]_i-\bar{\hat{\underline{a}}}_{\mathrm{J}})}{\sigma_{\mathrm{I}} \sigma_{\mathrm{J}}}
%\end{equation}
where $\underline{\hat{a}}_\mathrm{I}$ denotes the vectorization of $\underline{\hat{A}}_\mathrm{I}$. 
%Moreover,
%$$
%\textstyle{
%\sigma_{\mathrm{I}} := \sqrt{\sum_{i=1}^{n(n-1)}([\underline{\hat{a}}_\mathrm{I}]_i-\bar{\hat{\underline{a}}}_\mathrm{I})^2}, \qquad \bar{\hat{\underline{a}}}_\mathrm{I}:=\frac{\sum_{i=1}^{n(n-1)} [\underline{\hat{a}}_\mathrm{I}]_i}{n(n-1)}
%}
%$$
%while analogous notation holds for $\sigma_\mathrm{J}$ and $\bar{\hat{\underline{a}}}_\mathrm{J}$. \\
For synthetic data the same dissimilarity measure can also be computed w.r.t. the ground-truth, i.e. 
$\rho_{EC}(\mathrm{I}):=\rho(\underline{a},\underline{\hat{a}}_\mathrm{I})$ between the true $A$ and the estimated $\hat{A}_\mathrm{I}$.

\subsubsection{Functional Connectivity Metrics.} 
FC between two brain regions is the Pearson's correlation coefficient between the corresponding BOLD time-series. Hence, the FC matrix computed from a dataset $\mathcal{D}$ containing $n$ time-series $Y_i=[y_i(1),...,y_i(N)]^T$, $i=1,...,n$, is given by
\begin{equation}\label{equ:fc}
[F(\mathcal{D})]_{ij} := \rho(Y_i,Y_j), \qquad i,j=1,...,n
\end{equation}
with $\rho(\cdot,\cdot)$ as defined in Eq.~\eqref{equ:rho}.
%\begin{equation}\label{equ:fc}
%[F(\mathcal{D})]_{ij} := \frac{\sum_{k=1}^N \left(y_i(k)-\bar{y}_i\right)\left(y_j(k)-\bar{y}_j\right)}{\sqrt{\sum_{k=1}^N \left(y_i(k)-\bar{y}_i\right)^2}\sqrt{\sum_{k=1}^N \left(y_j(k)-\bar{y}_j\right)^2}}
%\end{equation}
%where $\bar{y}_i =\frac{1}{N}\sum_{k=1}^N y_i(k)$, $i=1,...,n$. \\
We also consider the FC matrix $\widehat{F}_\mathrm{I}$ obtained from the output  correlations of the estimated models, where the EC has been estimated 
according to modeling assumption $\mathrm{I}\in\mathcal{I}$, namely 
\begin{equation}
[\widehat{F}_{\mathrm{I}}]_{ij} := \frac{[\widehat{\Sigma}_{y}]_{ij}}{\sqrt{[\widehat{\Sigma}_{y}]_{ii}\ [\widehat{\Sigma}_{y}]_{jj}}}, \qquad i,j=1,...,n
\end{equation}
%where 
% To this purpose models \eqref{equ:linear_stochastic_dcm} or \eqref{equ:linear_stochastic_dcm_ext} are simulated using the inferred parameters $\hat{\theta}$, thus producing the BOLD time-series $\widehat{Y}_i=[\hat{y}_i(1),...,\hat{y}_i(N)]^T$, $i=1,..,n$. These are then used to compute $\widehat{F}_\mathrm{I}$ according to Eq.~\eqref{equ:fc}. However, for small values of $N$ (such as $N\sim 300$), different noise realizations lead to significantly different $\widehat{F}_\mathrm{I}$. To overcome such issue, we exploit the asymptotic stationarity of processes $\mathbf{x}(k)$ and $y(k)$ in Eq.~\eqref{equ:linear_stochastic_dcm} and \eqref{equ:linear_stochastic_dcm_ext} to compute the asymptotic FC matrix, say $\widehat{F}_\mathrm{I}^\infty$, starting from the estimated parameters $\hat{\theta}$. Namely,
%\begin{equation}
%[\widehat{F}_{\mathrm{I}}^\infty]_{ij} := \frac{[\widehat{\Sigma}^{\infty}_{y}]_{ij}}{\sqrt{[\widehat{\Sigma}^{\infty}_{y}]_{ii}\ [\widehat{\Sigma}^{\infty}_{y}]_{jj}}}
%\end{equation}
where $\widehat{\Sigma}_{y} = \widehat{\mathbf{C}}\widehat{\Sigma}_\mathbf{x}  \widehat{\mathbf{C}}^T + \hat{\epsilon} I_n$ and $\widehat{\Sigma}_\mathbf{x} $ is the solution of the Lyapunov equation $\widehat{\Sigma}_\mathbf{x} = \widehat{\mathbf{A}}\widehat{\Sigma}_{\mathbf{x}} \widehat{\mathbf{A}}^T + \widehat{\mathbf{Q}}_w$ if model \eqref{equ:linear_stochastic_dcm} is used. Alternatively, if model \eqref{equ:linear_stochastic_dcm_ext} is adopted, $\widehat{\Sigma}_\mathbf{x} $ is the solution of $\widehat{\Sigma}_{\mathbf{x}} = \widehat{\mathbf{M}}\widehat{\Sigma}_{\mathbf{x}} \widehat{\mathbf{M}}^T + \widehat{\mathbf{Q}}_\eta$.\\
To assess  how well the estimated EC (and thus the model), under  the modeling assumption $\mathrm{I}\in\mathcal{I}$, is able to capture the empirical FC, we evaluate the following Pearson's correlation coefficient 
\begin{equation}
\rho_{\mathrm{FC}}(\mathcal{D},\mathrm{I}) := \rho(f(\mathcal{D}), \hat{f}_\mathrm{I}),
\end{equation}
where $ \hat{f}_\mathrm{I}$ and $f(\mathcal{D})$ respectively denote the vectorization of the upper diagonal parts of $\widehat{F}_{\mathrm{I}}$ and of the FC matrix computed with the estimation data $\mathcal{D}$. \\
When dealing with the  synthetic experiment we can also  use the test dataset $\mathcal{D}^{te}$ to compute $\rho_\mathrm{FC}(\mathcal{D}^{te},\mathrm{I}):= \rho(f(\mathcal{D}^{te}), \hat{f}_\mathrm{I})$.

\begin{figure}[h]
\centering
\subfigure[White noise data]{\includegraphics[scale=.315]{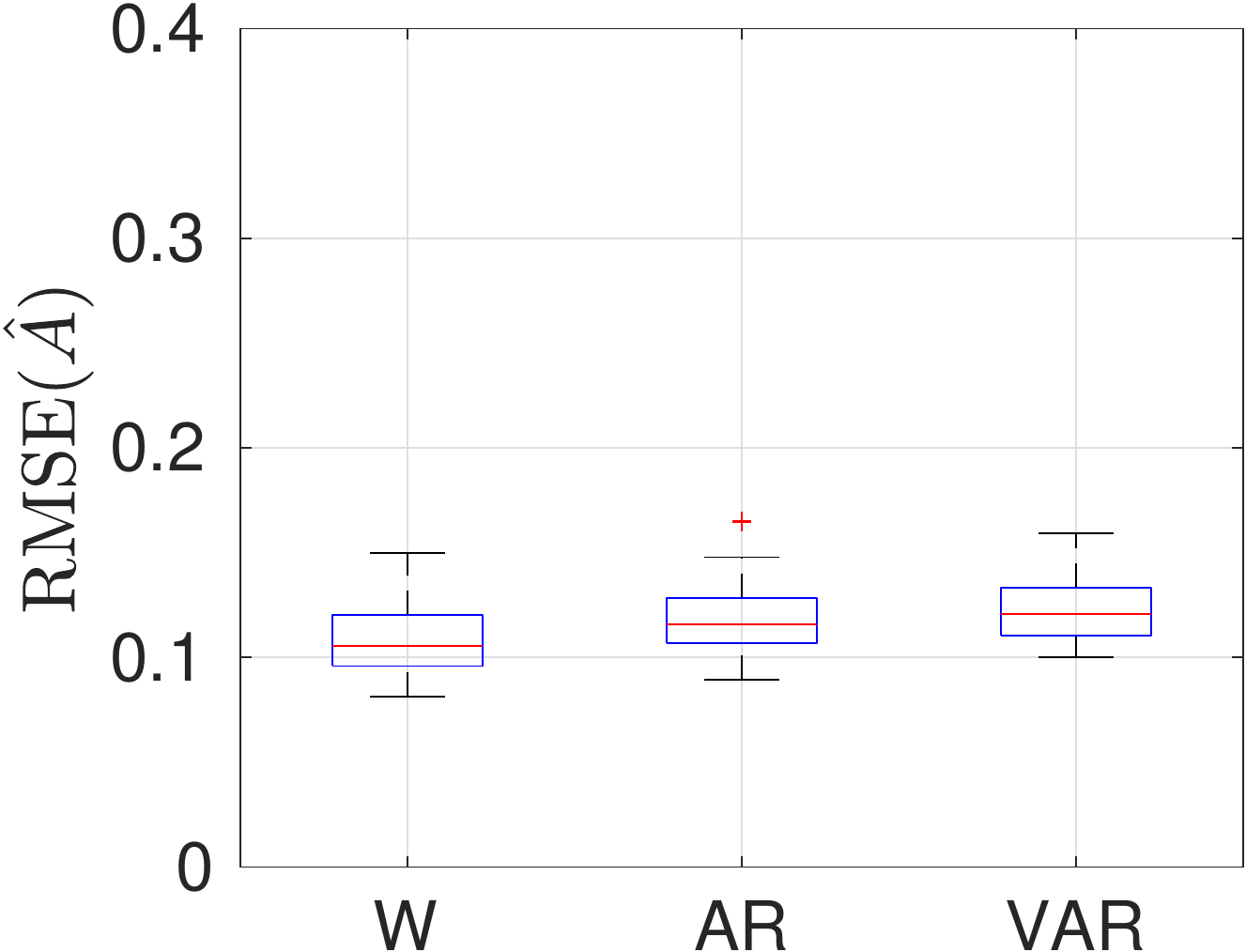}}
\quad
\subfigure[VAR noise data]{\includegraphics[scale=.31]{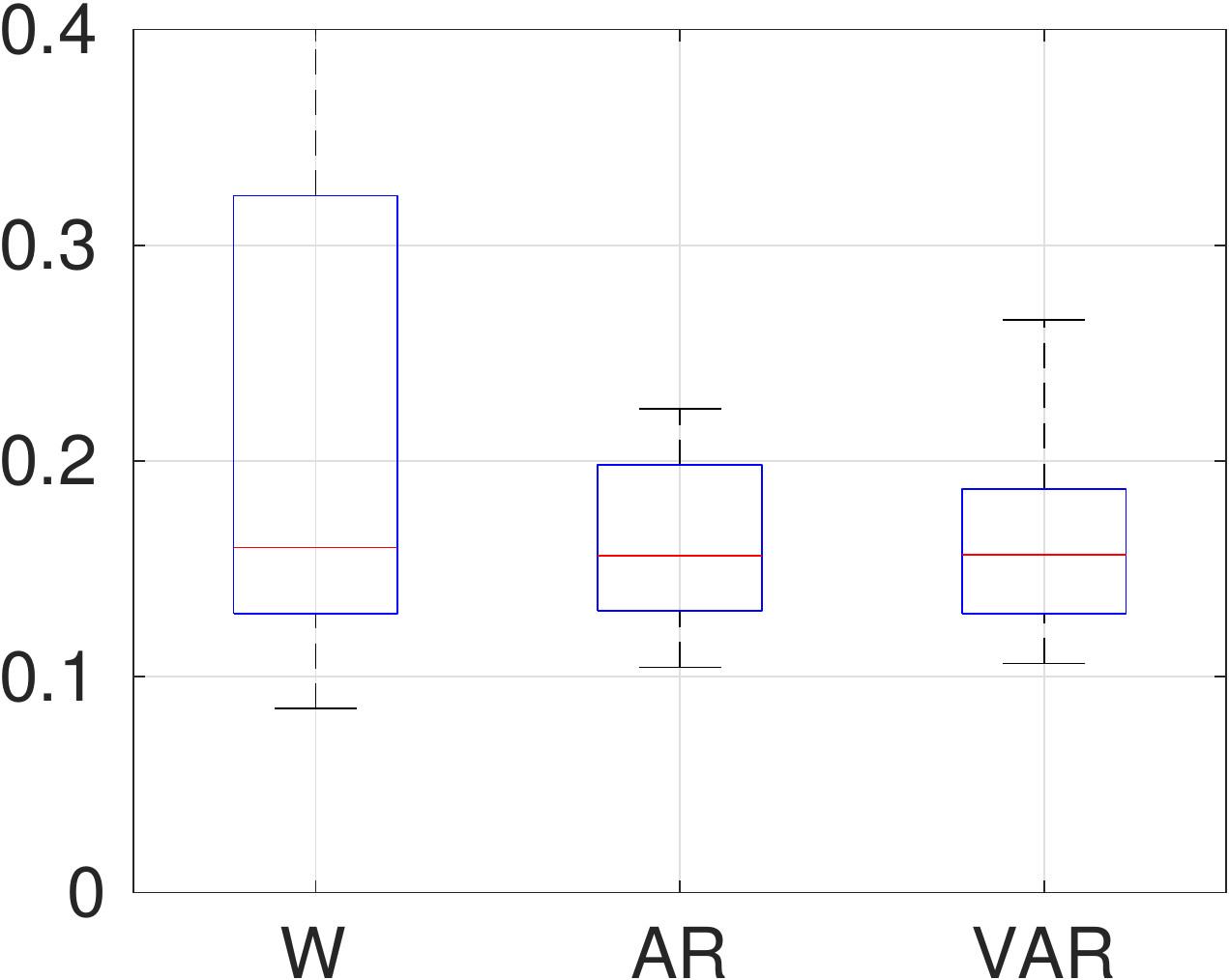}}
\caption{Synthetic data: boxplot of RMSE$(\hat{A}_{\mathrm{I}})$, $\mathrm{I}\in\{\mathrm{W,AR,VAR}\}$ over 50 Monte-Carlo runs.}\label{fig:rmse}
\end{figure}

\begin{figure}[h]
\centering
\subfigure[White noise data]{\includegraphics[scale=.315]{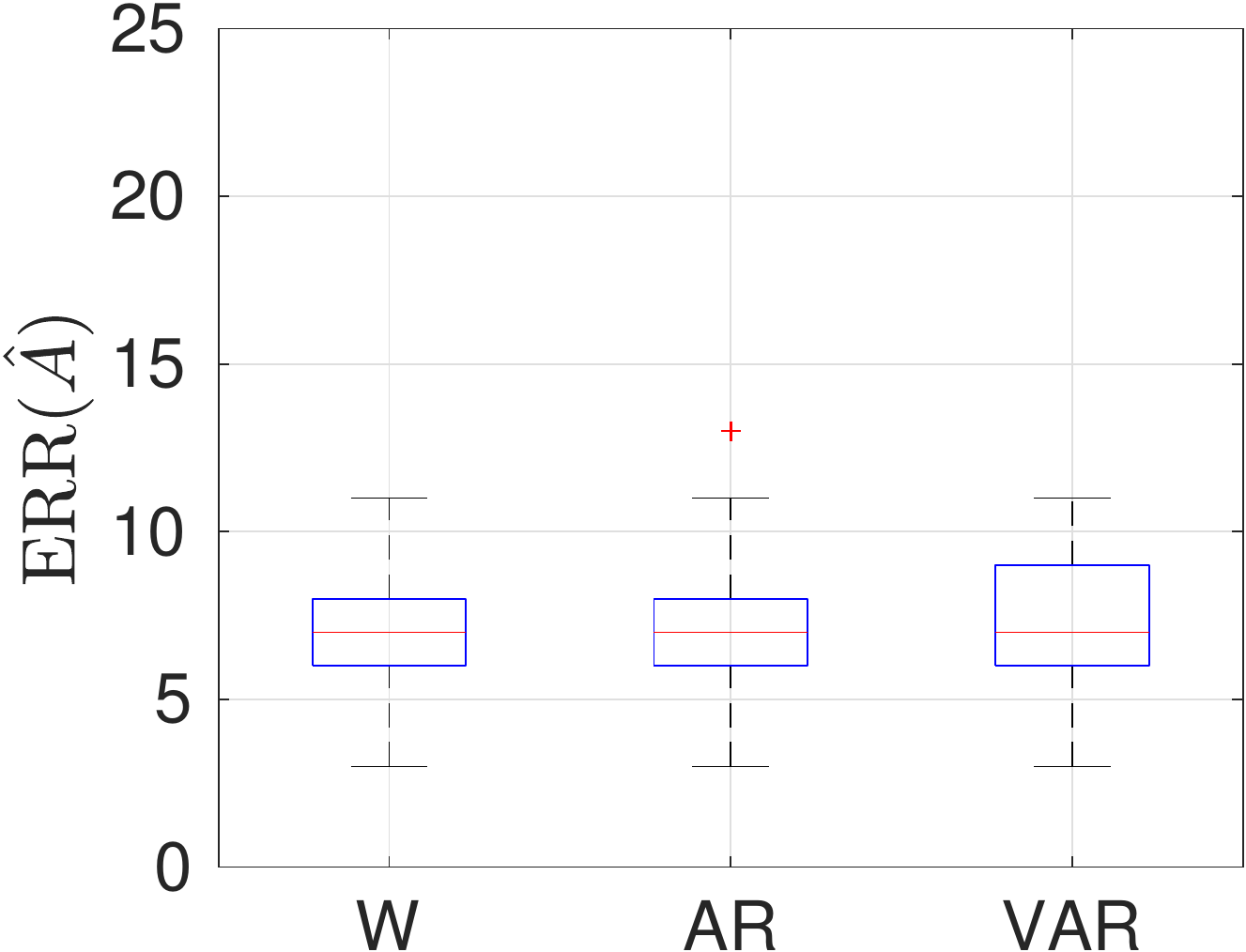}}
\quad
\subfigure[VAR noise data]{\includegraphics[scale=.31]{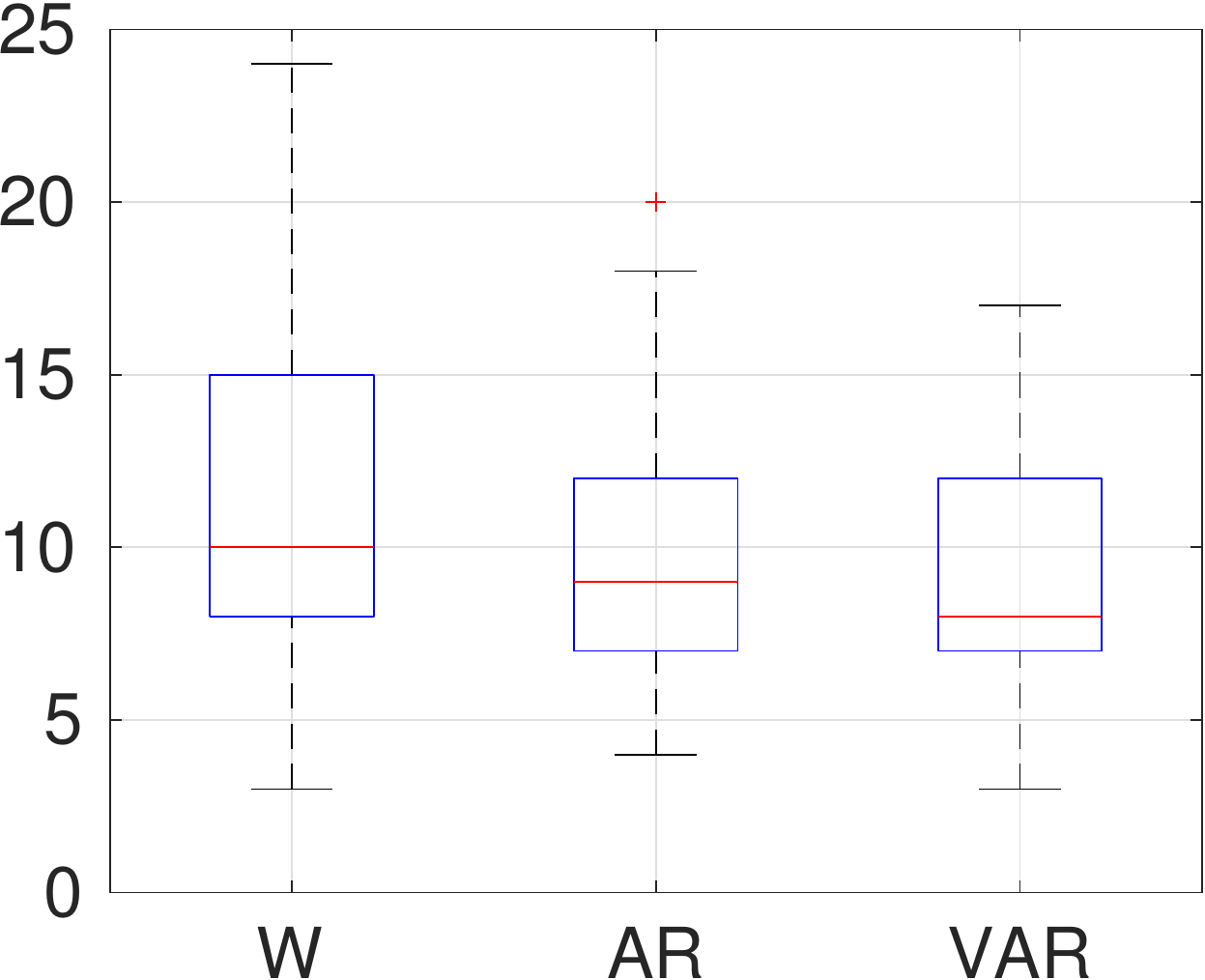}}
\caption{Synthetic data: boxplot of ERR$(\hat{A}_{\mathrm{I}})$, $\mathrm{I}\in\{\mathrm{W,AR,VAR}\}$ over 50 Monte-Carlo runs.}\label{fig:err}
\end{figure}

\begin{figure}[h]
\centering
\subfigure[White noise data]{\includegraphics[scale=.305]{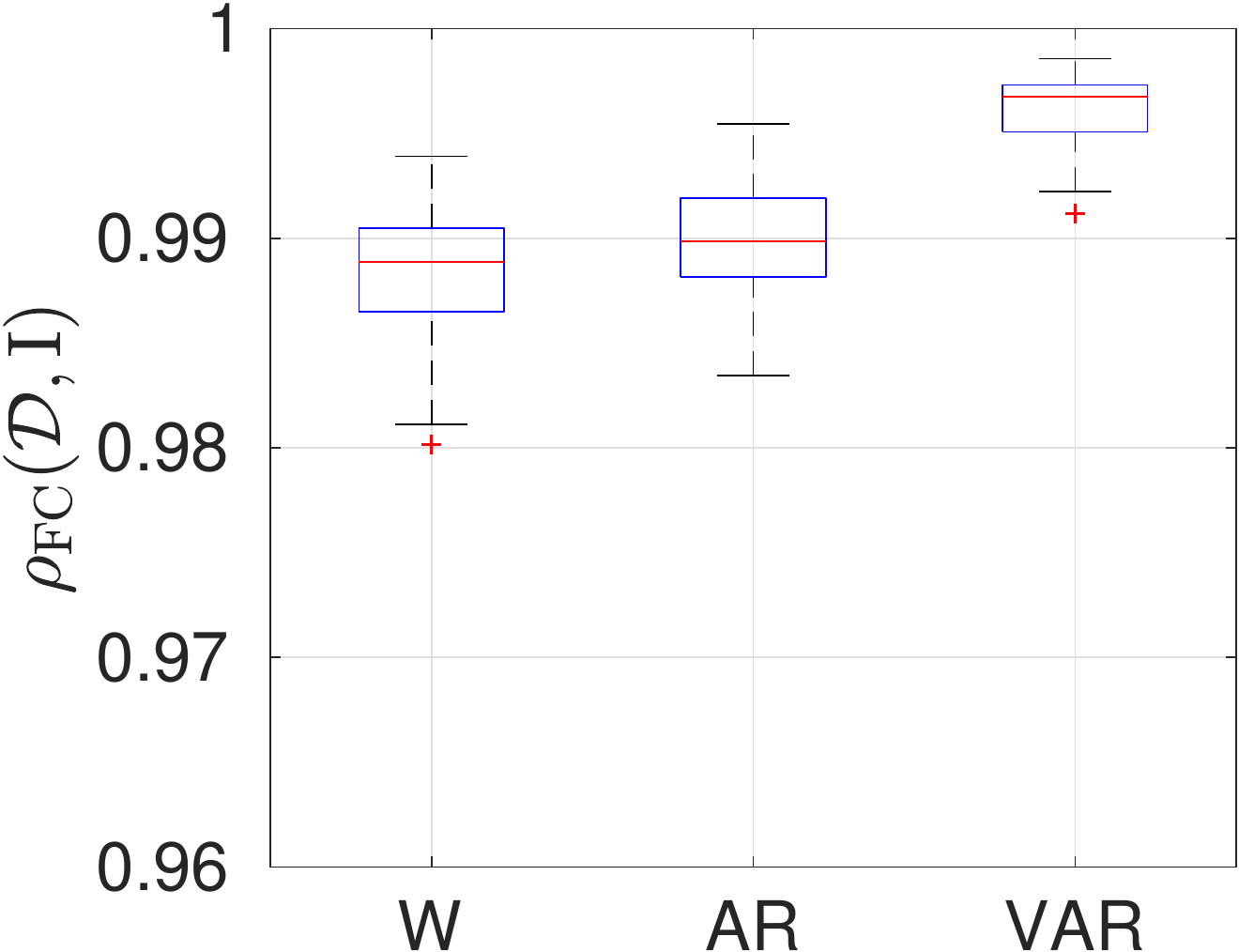}}
\quad
\subfigure[VAR noise data]{\includegraphics[scale=.30]{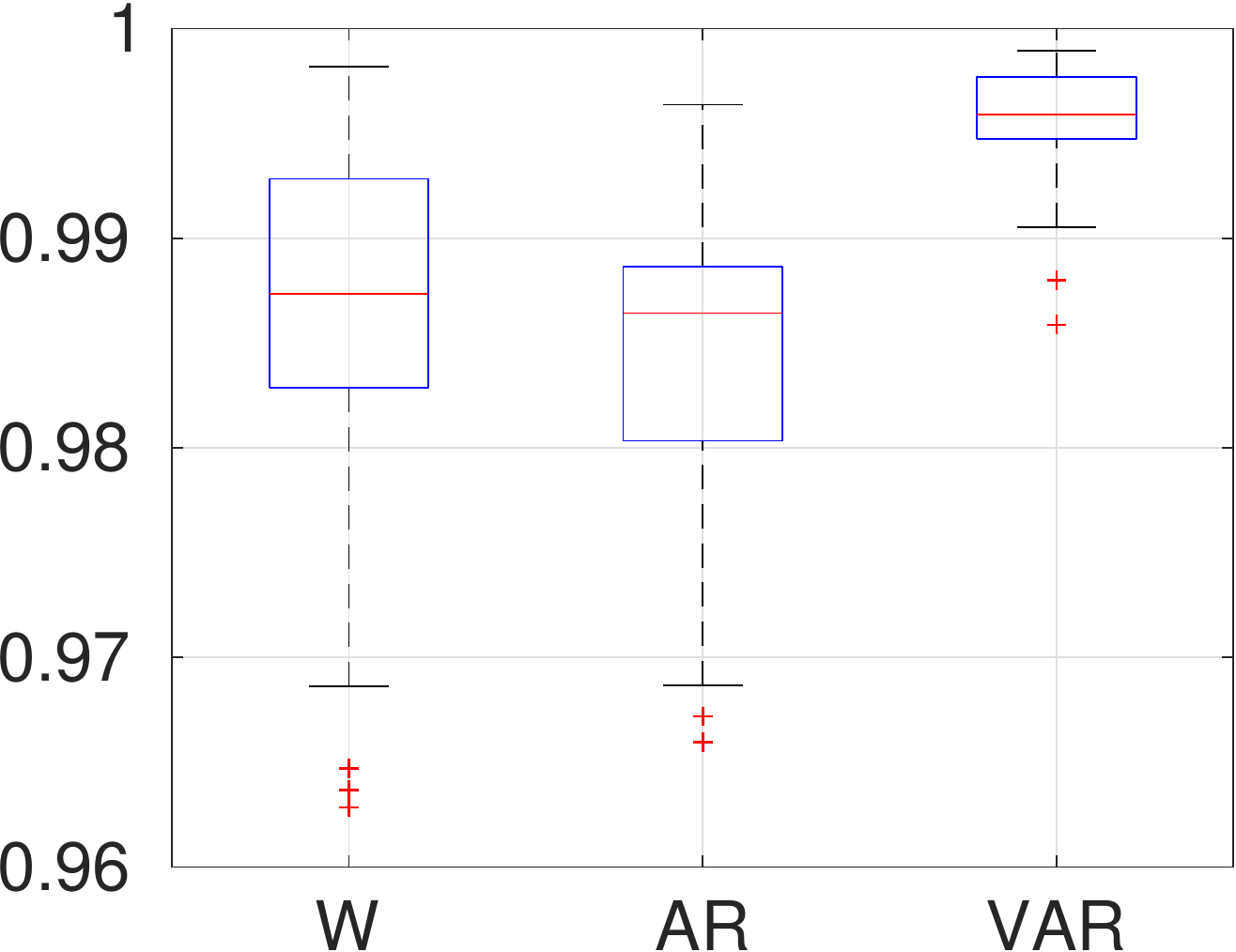}}
\caption{Synthetic data: boxplot of $\rho_{\mathrm{FC}}(\mathcal{D}_j,\mathrm{I}), \ j=1,..,50,\ \mathrm{I}\in\{\mathrm{W,AR,VAR}\}$.}\label{fig:corrFC}
\end{figure}

\begin{figure}[h]
\centering
\subfigure[White noise data]{\includegraphics[scale=.305]{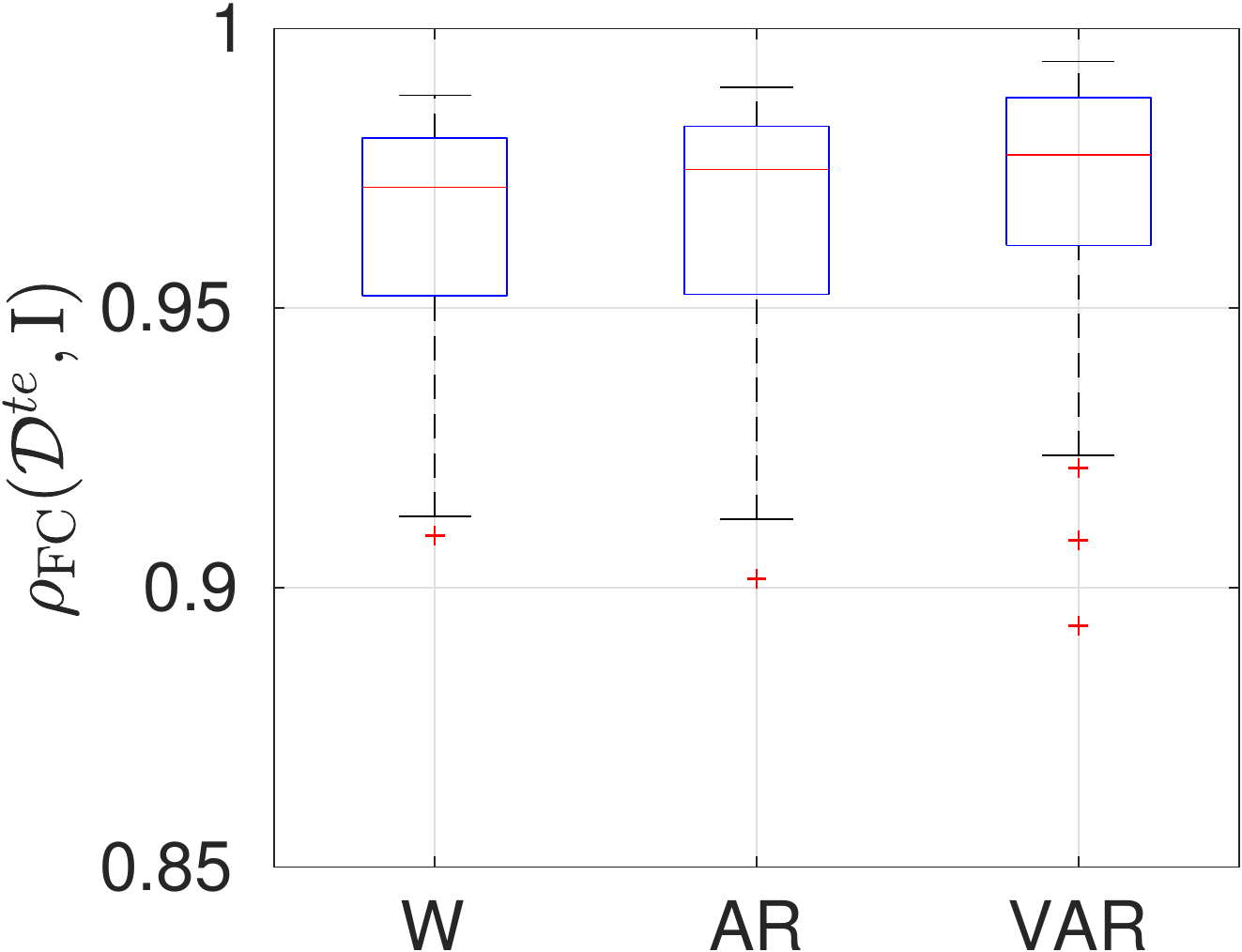}}
\quad
\subfigure[AR noise data]{\includegraphics[scale=.30]{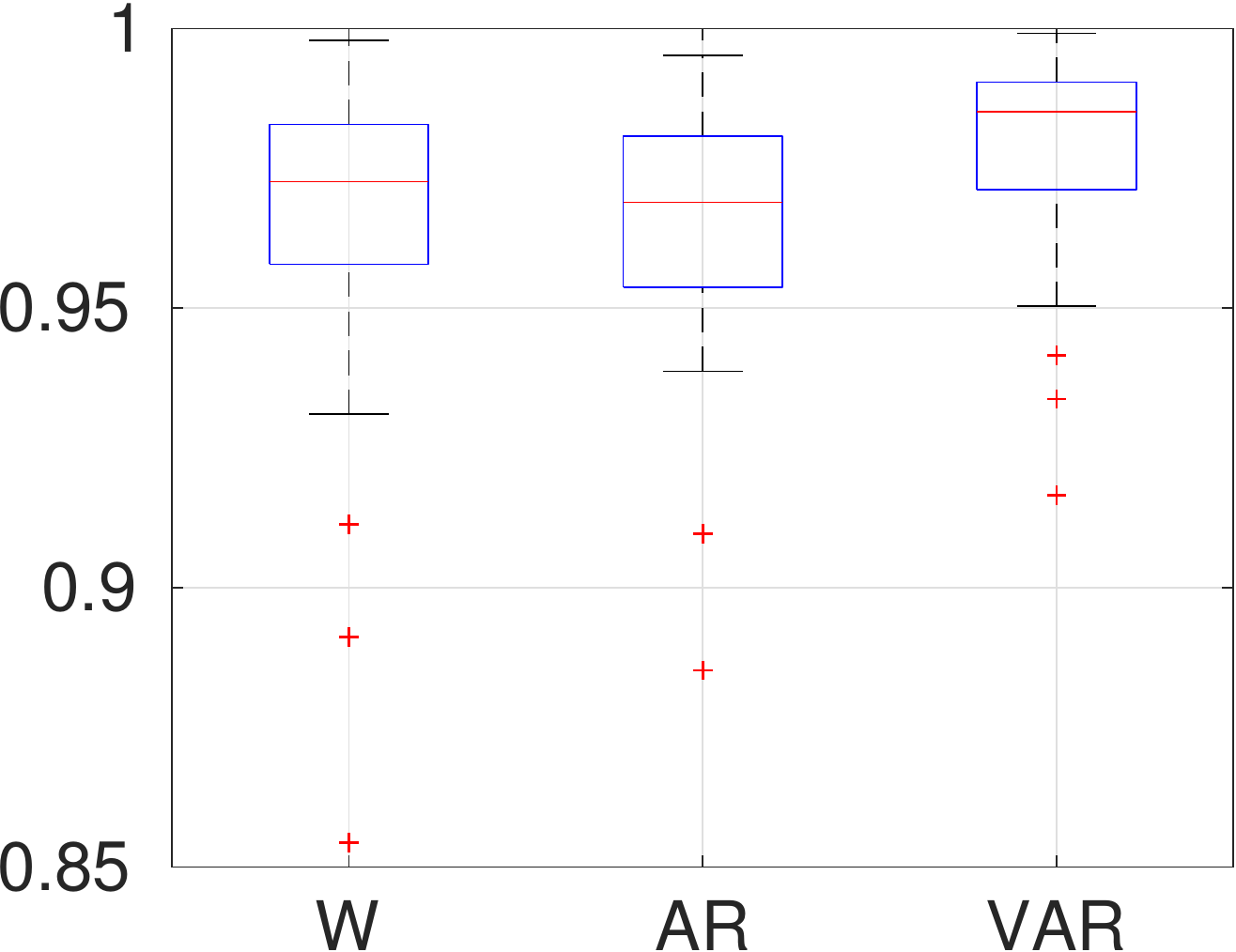}}
\caption{Synthetic data: boxplot of $\rho_{\mathrm{FC}}({D}_j^{te},\mathrm{I}),\ j=1,..,50,\ \mathrm{I}\in\{\mathrm{W,AR,VAR}\}$.}\label{fig:corrFCtest}
\end{figure}

\begin{figure}[h]
\centering
\subfigure[White noise data]{\includegraphics[scale=.42]{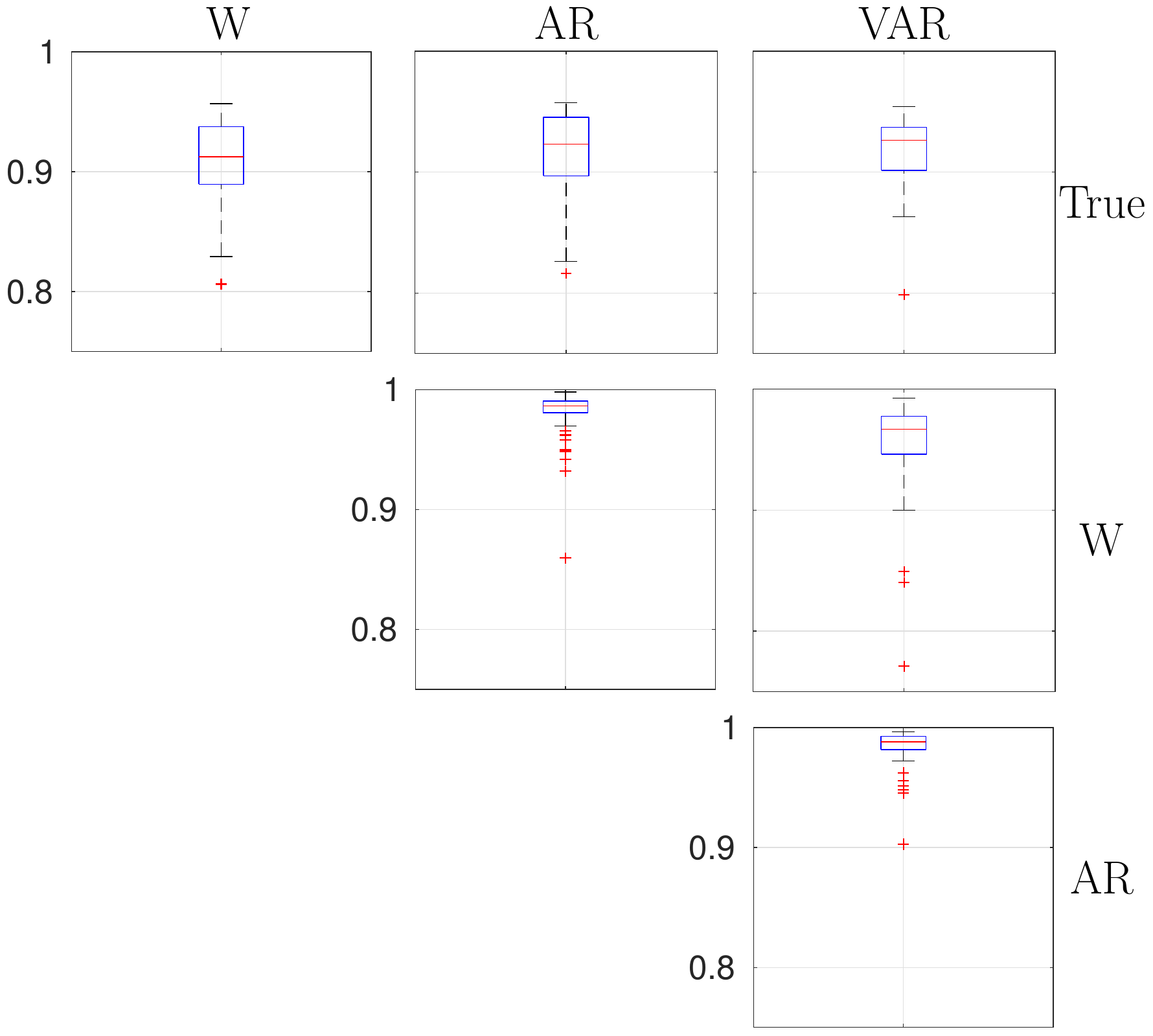}}
\quad
\subfigure[VAR noise data]{\includegraphics[scale=.42]{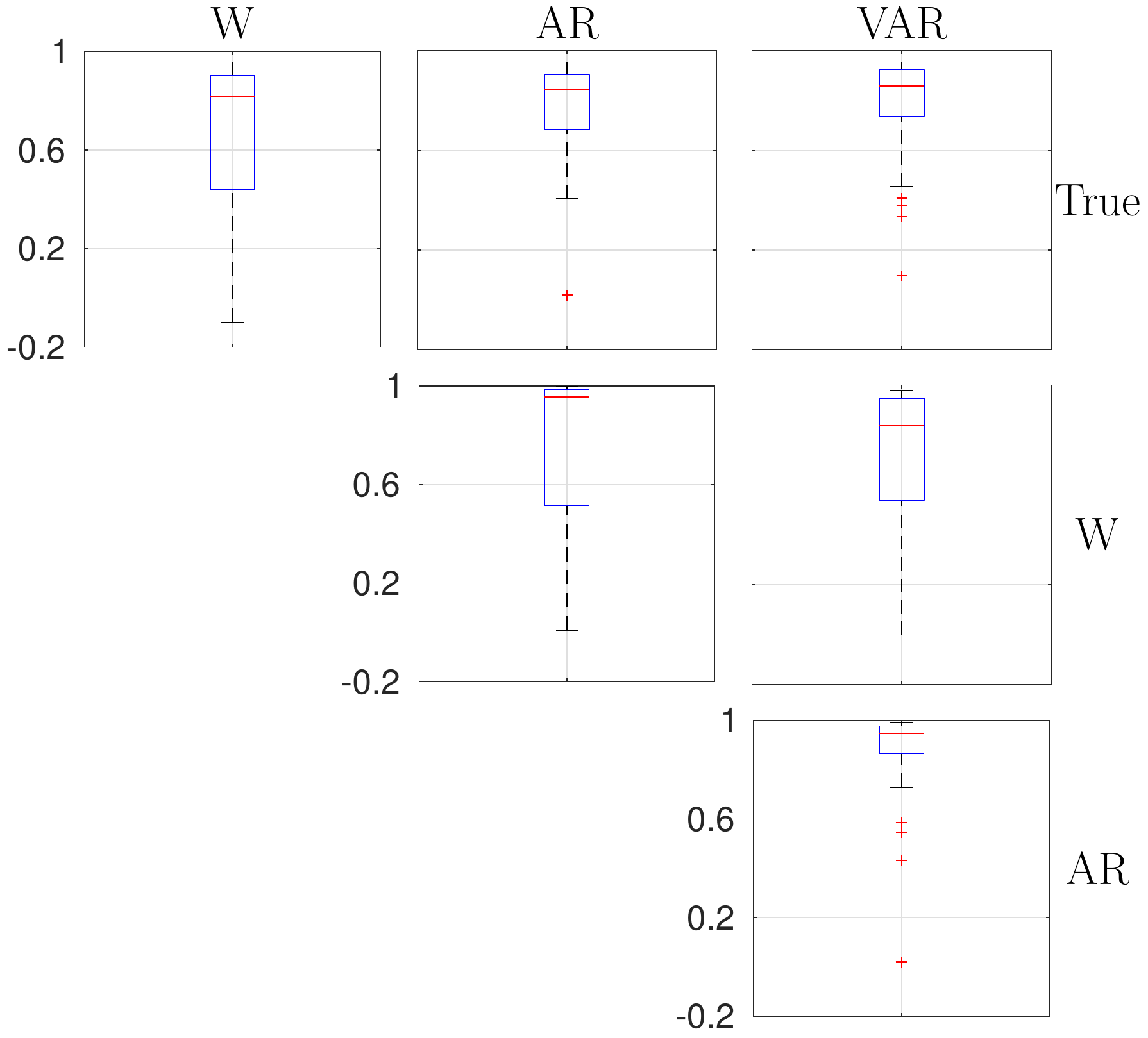}}
\caption{Synthetic data: boxplot of $\rho_\mathrm{EC}(\mathrm{I},\mathrm{J})$, $\mathrm{I,J}\in\{\mathrm{W,AR,VAR}\}$.}\label{fig:corrEC}
\end{figure}

\subsection{Results}\label{sec:results}
\subsubsection{Synthetic Data.}
Figs.~\ref{fig:rmse} and \ref{fig:err} show the results obtained in the synthetic datasets in terms of metrics \eqref{equ:rmse} and \eqref{equ:err}, respectively.  The three modeling assumptions $\mathcal{I}=\{\mathrm{W,AR,VAR} \}$ give rise to similar performance, independently of how the data have been generated. The unique exception is observed on the data generated from VAR process noise when a white noise is postulated for the brain endogenous fluctuations $w(t)$: the RMSE is significantly larger than that obtained assuming a colored process noise (that is, under assumptions AR and VAR). However, such poor performance does not negatively impact the metrics based on FC, that is $\rho_{\mathrm{FC}}(\mathcal{D},\mathrm{I})$ and $\rho_{\mathrm{FC}}(\mathcal{D}^{te},\mathrm{I})$, as can be noticed in Figs.~\ref{fig:corrFC}-\ref{fig:corrFCtest}. The similarity of the performance produced by the three modeling assumptions is particularly evident in Fig.~\ref{fig:corrFCtest}, which reports the correlation between the output FC of the estimated model and the empirical FC computed on test dataset. These outcomes show that all the estimated models are equally able to reproduce both estimation and test data, even if the EC matrix is not correctly inferred, thus suggesting a possible identifiability problem. Furthermore, these results seem to discredit the reliability of metrics $\rho_{\mathrm{FC}}(\mathcal{D}, \mathrm{I})$ in assessing the goodness of the estimated EC. However, such performance index is commonly used for this purpose when a ground truth is not available, that is when real data are treated.\\
If we consider the correlation among the estimated ECs the situation is quite different: when data are generated by a white process noise (Fig.~\ref{fig:corrEC}(a)), $\rho_\mathrm{EC}(\mathrm{I},\mathrm{J})$ takes values which are very close to 1, regardless of the modeling assumption. On the other hand, when a VAR process is responsible for the excitation of model \eqref{equ:linear_stochastic_dcm_ext} (Fig.~\ref{fig:corrEC}(b)), the EC estimated under hypothesis $\mathrm{W}$ is much less similar to those obtained from hypothesis $\mathrm{AR}$ and $\mathrm{VAR}$. In this case $\rho_{\mathrm{EC}}(\mathrm{AR},\mathrm{VAR})$ is still large, but smaller than what is reported in Fig.~\ref{fig:corrEC}(a). The correlations $\rho_{\mathrm{EC}}(\mathrm{I})$ between the true EC matrix $A$ and the estimated ones (first row of plots in Fig.~\ref{fig:corrEC}) reflect what we observe in Figs.~\ref{fig:rmse} and \ref{fig:err}: while $\rho_\mathrm{EC}(\mathrm{I})$, $\mathrm{I}\in\mathcal{I}$, are very similar in Fig.~\ref{fig:corrEC}(a), in Fig.~\ref{fig:corrEC}(b) the ECs estimated postulating a white process noise appear quite dissimilar from the true one, w.r.t. to those returned by $\mathrm{AR}$ and $\mathrm{VAR}$ modeling assumptions.

\subsubsection{Real Data.} 
\begin{figure}[h]
\centering
\subfigure[Data $\mathcal{D}^{\mathrm{VIS}}$]{\includegraphics[scale=.82]{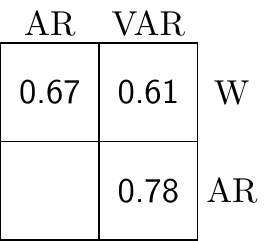}}
\quad
\subfigure[Data $\mathcal{D}^{\mathrm{DMN}}$]{\includegraphics[scale=.82]{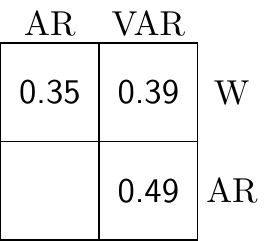}} 
\quad
\subfigure[Data $\mathcal{D}^{\mathrm{wb}}$]{\includegraphics[scale=.82]{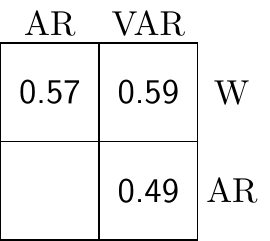}} 
\caption{Real data: $\rho_\mathrm{EC}(\mathrm{I},\mathrm{J})$, $\mathrm{I,J}\in\{\mathrm{W,AR,VAR}\}$.}\label{tab:corrEC}
\end{figure}

Considering  real data, we have to rely only on the metrics $\rho_{\mathrm{FC}}(\mathcal{D},\mathrm{I})$ and $\rho_\mathrm{EC}(\mathrm{I},\mathrm{J})$  since a ground truth is not available. Table~\ref{tab:corrFC} reports the correlations between the FCs ($F(\mathcal{D})$) computed from the BOLD time-series and the model FCs ($\widehat{F}_\mathrm{I}$)  produced by the systems estimated under the modeling assumptions $\{\mathrm{W,AR,VAR}\}$. Not surprisingly, in all the three datasets, these hypothesis give rise to equal $\rho_{FC}(\mathcal{D},\mathrm{I})$ values, in agreement with the results obtained  on synthetic data. However, if we inspect the correlation $\rho_\mathrm{EC}(\mathrm{I},\mathrm{J})$ between the estimated ECs (Fig.~\ref{tab:corrEC}), the situation more closely resembles what reported in Fig.~\ref{fig:corrEC}(b) for synthetic data. Indeed, the inferred ECs result to be quite different, according to the similarity metrics we adopt. This could suggest that modeling the brain endogenous fluctuations as a VAR process may be a more realistic hypothesis. However, recalling that the synthetic scenario considers a network of 7 brain regions, while the simulations on real data refer to  $\sim$50 regions, we should stress that the previous statement requires further numerical experiments, as well as sound statistical analysis, to be validated. Moreover, we should observe how $\rho_\mathrm{EC}(\mathrm{AR},\mathrm{VAR})$ takes much smaller values in the real data scenario,  w.r.t. what reported in the synthetic setup. Hence, further analyses are required also to understand the reasons of this disagreement.

\begin{table}[h]
\begin{center}
\begin{tabular}{|l|ccc|}
\hline
Dataset & \multicolumn{3}{c|}{Modeling Assumption}\\
& W & AR & VAR \\
\hline
$\mathcal{D}^{\mathrm{VIS}}$ & 0.93 & 0.92 & 0.92\\
$\mathcal{D}^{\mathrm{DMN}}$ & 0.94 & 0.95 &0.94\\
$\mathcal{D}^{wb}$ & 0.92 & 0.92 & 0.92 \\
\hline 
\end{tabular}
\end{center}
\vspace{2mm}

\caption{Real data: $\rho_\mathrm{FC}(\mathcal{D},\mathrm{I})$,\hspace{2cm} $\mathcal{D}\in\{\mathcal{D}^{\mathrm{VIS}},\mathcal{D}^{\mathrm{DMN}},\mathcal{D}^{wb}\}$, $\mathrm{I}\in\{\mathrm{W,AR,VAR}\}$.}\label{tab:corrFC}
\end{table}

%\begin{table}[h]
%\centering
%\begin{tabular}{|l|cc|}
%\hline
%& AR & VAR \\
%\hline
%W  & 0.67 & 0.61 \\
%AR  & & 0.78  \\
%\hline 
%\end{tabular}
%
%\vspace{2mm}
% 
%\caption{Dataset $\mathcal{D}^{\mathrm{VIS}}$: $\rho_\mathrm{EC}(\mathrm{I},\mathrm{J})$, $\mathrm{I,J}\in\{\mathrm{W,AR,VAR}\}$.}\label{tab:corrEC_dataVIS}
%\end{table}

\section{Conclusions}\label{sec:conclusions}
In this work we have tackled the estimation of brain effective connectivity from both synthetic and real resting-state fMRI data. Following our previous work \citep{prando2017estimating}, we describe the dynamics of the neuronal activity by means of a linear state-space model, which also includes a linearized hemodynamic response, in charge of mapping the neuronal activity to the so-called BOLD signal, measured by fMRI. At rest, brain activity is elicited by endogenous fluctuations, which play the role of process noise in the aforementioned state-space system. While in \cite{prando2017estimating}, this was considered white, here we reformulate our generative model in order to account for the presence of colored process noise. Specifically, we assume it to be a 1st-order autoregressive process. Such hypothesis is motivated by previous works in the neuroscience community, where brain fluctuations are characterized as low-frequency signals \citep{friston2014dcm,linkenkaer2001long}.\\
Not surprisingly, using an AR process noise in the generative model seems a more robust choice, independently of the noise type which has actually generated the data. Furthermore, despite its widespread use in neuroscience, we argue that functional connectivity does not allow to discriminate between the modeling assumptions of white or AR process noise. This is to be expected since FC is just a zero lag output correlation, which is largely insufficient to identify the model parameters. \\
Finally, evaluating the correlation among the ECs estimated under different modeling assumptions, we find a partial agreement between the results on the real data and those obtained on synthetic data generated by an AR process noise. However, we believe that deeper investigations are required to validate such statement. To this purpose we are conducting bootstrap simulations to compare synthetic and real experiments on solid statistical grounds.

\bibliography{References_SYSID18}             % bib file to produce the bibliography
                                                     % with bibtex (preferred)

\end{document}